\documentclass[aps,reprint,superscriptaddress,longbibliography]{revtex4-1}
\usepackage{times,graphicx,jab,color}

\begin{document}
\title{Evidence for Electron Landau Damping in Space Plasma Turbulence}
\date{\today}
\author{C.~H.~K.~Chen*}
\affiliation{School of Physics and Astronomy, Queen Mary University of London, London E1 4NS, UK}
\author{K.~G.~Klein}
\affiliation{Lunar and Planetary Laboratory, University of Arizona, Tucson, AZ 85719, USA}
\author{G.~G.~Howes}
\affiliation{Department of Physics and Astronomy, University of Iowa, Iowa City, IA 52242, USA}
\begin{abstract}
\section*{Abstract} 
How turbulent energy is dissipated in weakly collisional space and astrophysical plasmas is a major open question. Here, we present the application of a field-particle correlation technique to directly measure the transfer of energy between the turbulent electromagnetic field and electrons in the Earth's magnetosheath, the region of solar wind downstream of the Earth's bow shock. The measurement of the secular energy transfer from the parallel electric field as a function of electron velocity shows a signature consistent with Landau damping. This signature is coherent over time, close to the predicted resonant velocity, similar to that seen in kinetic Alfv\'en turbulence simulations, and disappears under phase randomisation. This suggests that electron Landau damping could play a significant role in turbulent plasma heating, and that the technique is a valuable tool for determining the particle energisation processes operating in space and astrophysical plasmas.
\end{abstract}
\maketitle

\section*{Introduction}

It is well-established that turbulence pervades space and astrophysical plasmas, transferring energy from the large scales at which it is injected down to the plasma microscales where it can be dissipated. The resulting plasma heating is thought to be dynamically important in a number of systems, e.g., the solar corona and solar wind \citep{cranmer15}, the interstellar medium \citep{scalo04}, and galaxy clusters \citep{zhuravleva14}, although it is not yet known which physical dissipation mechanisms are responsible. It is therefore a major open question as to how turbulent plasma heating occurs, although due to the weakly collisional nature of these plasmas, it is inevitably through a series of different microphysical plasma processes. In this paper, we apply a field-particle correlation technique to \emph{in situ} spacecraft data to investigate the first step in the thermalisation process: the mechanism by which energy is transferred from the turbulent electromagnetic field to the plasma particles.

The solar wind provides an ideal opportunity to study turbulent heating, due to the high-resolution \emph{in situ} measurements available, and several different mechanisms have been proposed. Early suggestions \citep{coleman68} invoked  cyclotron damping to enable perpendicular ion energisation \citep{marsch82c,isenberg83,cranmer14}. The realisation that the turbulence could have a substantial $k_\perp$ component led also to suggestions of Landau damping \citep{dobrowolny85,leamon99} and later work predicted that this would be dominant over cyclotron damping \citep{quataert98,howes08a,schekochihin09} due to the anisotropic nature of the turbulent cascade \citep{horbury12,oughton15,chen16b}. Many models now incorporate the effect of both ion and electron Landau damping \citep{sahraoui09,howes11c,borovsky11b,tenbarge13b,passot15}, although recent work has raised interesting questions about how effective this is in turbulent systems \citep{plunk13,parker16,schekochihin16,verscharen17}. Non-resonant mechanisms have also been proposed, most notably stochastic heating \citep{johnson01,voitenko04,chandran10b}, which leads to the broadening of particle distributions in a stochastic field. It has also been suggested that dissipation is localized at structures, such as reconnecting current sheets \citep{retino07,sundkvist07}, vortices \citep{alexandrova06,haynes15}, and double layers \citep{stawarz15}, although the question remains which dissipation processes would occur within such structures \citep{egedal12,tenbarge13a,loureiro13,drake14,chen15,howes18}.

Various observational evidence has been presented for the above mechanisms, although to date this has been somewhat indirect. For example, evidence for cyclotron damping has been based on the wavenumber of the ion-scale break in the turbulence spectrum \citep{coleman68,denskat83,leamon98a,bourouaine12,bruno14a,woodham18}, the shape of contours in the ion distributions \citep{marsch01,he15a}, or correlations between species temperatures and drifts \citep{gary05,kasper08,kasper13}. Similarly, evidence for stochastic heating has been based on relationships between measured temperatures and turbulence amplitudes \citep{bourouaine13,chandran13,vech17}. Localised enhancements in temperature \citep{retino07,osman12a,chasapis18short} and work done on the particles \citep{retino07,sundkvist07,chasapis18short} have also been cited as evidence for dissipation at structures.

In this paper, we present a direct measurement of the secular transfer of energy from the turbulent electromagnetic field at kinetic scales to the electrons as a function of the electron velocity. This velocity-space signature allows the different heating mechanisms to be identified, and here is found to be consistent with electron Landau damping.

\section*{Results}

\textbf{Data set.} Data from the \emph{Magnetospheric Multiscale} (\emph{MMS}) mission \citep{burch16} were used, when the spacecraft were in the Earth's magnetosheath on 16th October 2015 09:24:11--09:25:21. The mean plasma parameters at this time were: magnetic field strength $B\approx 39$\,nT, number density $n_\mathrm{i}\approx n_\mathrm{e}\approx14$\,cm$^{-3}$, bulk velocity $u_\mathrm{i}\approx u_\mathrm{e}\approx180$\,km\,s$^{-1}$, and temperatures $T_\mathrm{\|i}\approx150$\,eV, $T_\mathrm{\perp i}\approx240$\,eV, $T_\mathrm{\|e}\approx22$\,eV, $T_\mathrm{\perp e}\approx23$\,eV. These correspond to average plasma betas $\beta_\mathrm{i}\approx0.80$ and $\beta_\mathrm{e}\approx0.088$ (where $\beta_s=2\mu_0n_sk_\mathrm{B}T_s/B^2$). Magnetic field data were measured by FGM \citep{russell16short} and SCM \citep{lecontel16short}, electric field data by SDP \citep{lindqvist16short} and ADP \citep{ergun16short}, and particle data by FPI \citep{pollock16short}. All data in this paper are from \emph{MMS3} and the turbulence measured during this time period was previously characterised \citep{chen17}.

\begin{figure}
\includegraphics[width=\columnwidth,trim=0 0 0 0,clip]{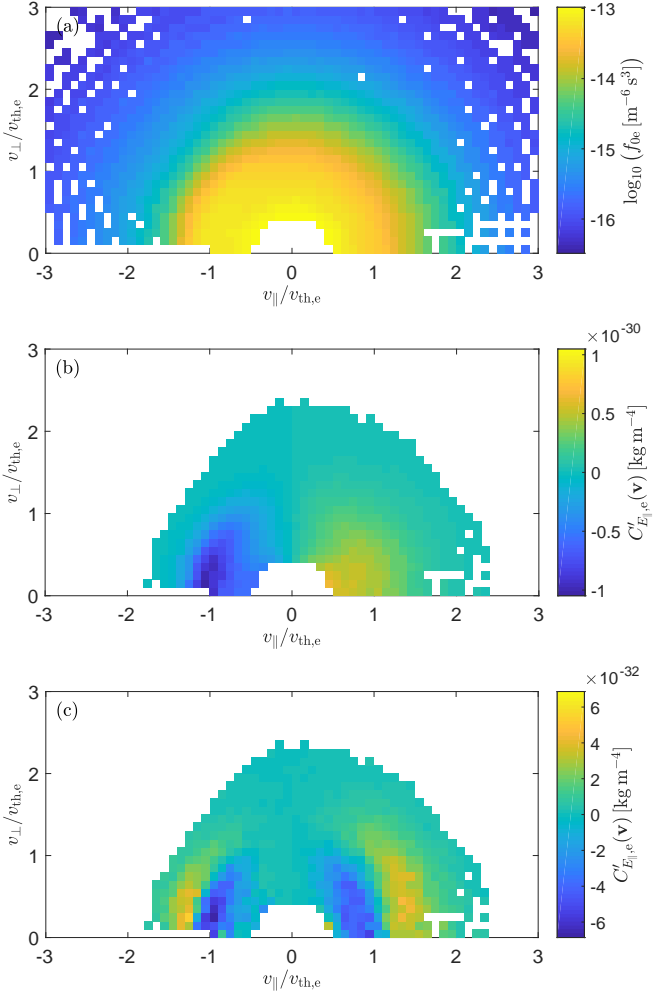}
\caption{\textbf{Measured average electron distribution and field-particle energy transfer rates.} (a) Average electron distribution $f_\mathrm{0e}$. (b) Alternative energy transfer rate $C'_{E_\|,\mathrm{e}}(\mathbf{v})$ using $f_\mathrm{e}$ and unfiltered $E_\|$. (c) Alternative energy transfer rate using $\delta f_\mathrm{e}$ and high-pass-filtered (at 1\,Hz) $E_\|$.}
\label{fig:2d}
\end{figure}

Here, we focus on the energy transfer to the electrons, which were measured at 30\,ms resolution, resulting in a total of 2,333 three-dimensional velocity distributions. The average of these, $f_\mathrm{0e}=\left<f_\mathrm{e}\right>$, is shown in Fig.~\ref{fig:2d}(a), in the frame in which the mean electron bulk flow is zero and in a coordinate system in which $v_\|$ is parallel to the global mean field $\mathbf{B}_0=\left<\mathbf{B}\right>$, $v_\perp=\sqrt{v_{\perp1}^2+v_{\perp2}^2}$, and $v_\mathrm{th,e}=\sqrt{2k_\mathrm{B}T_\mathrm{e}/m_\mathrm{e}}$ is the isotropic electron thermal speed. In the conversion from measured energy bin to particle velocity, the mean spacecraft potential (relative to the plasma) of $+4.2$\,V was subtracted to compensate for the energy gain of the electrons arriving at the positively charged spacecraft. Note that data is unavailable for the central part of the distribution with $v\lesssim0.5v_\mathrm{th,e}$.

\textbf{Measuring secular energy transfer.} The energy transfer was measured by calculating $C'_{E_\|,\mathrm{e}}(\mathbf{v})=\left<q_\mathrm{e}v_\|E_\|f_\mathrm{e}\right>$ [see Eq.~(\ref{eq:cepardash}) of the Methods section] at each point of the measured electron distributions, with the average taken over the whole interval. For the parallel electric field $E_\|$, the time series of electric field vectors (measured at $\approx0.12$\,ms resolution) was first Lorentz transformed to the zero mean bulk velocity frame \citep{chen11b}, averaged down to 30\,ms resolution, then the component parallel to $\mathbf{B}_0$ taken. \textcolor{black}{The $E_\|$ measurement here remains above the instrumental noise level for frequencies $\lesssim100$\,Hz, which covers the range used for the correlation measurement.} Since FPI was operating in interleave mode, in which alternate distributions were sampled at different points in velocity space \citep{pollock16short}, $C'_{E_\|,\mathrm{e}}(\mathbf{v})$ was calculated separately for each of the two sets of distributions. This results in an effective lower time resolution of 60\,ms (corresponding to a maximum wavenumber $k\rho_\mathrm{i}\approx34$, where $\rho_\mathrm{i}$ is the ion gyroradius, under the Taylor hypothesis) but greater coverage in velocity space when recombined. The resulting energy transfer measure, combined, binned, and averaged in $(v_\|,v_\perp)$ space, is shown in Fig.~\ref{fig:2d}(b). To ensure reliability, distribution measurements with fewer than 3 particle counts and greater than 20\% data gaps in time were excluded, leading to the reduced coverage.

Fig.~\ref{fig:2d}(b) shows a clear signature roughly antisymmetric about $v_\|=0$. However, this is likely due to the large-scale wave-like oscillation that dominates the energy transfer \citep{klein16b,howes17a,klein17a}. As discussed in the Methods section, part of the technique is to average out this oscillation to leave the secular transfer, however, in a turbulent spectrum, averaging over longer times leads to larger-scale oscillations dominating the transfer measurement. Instead, the $E_\|$ time series was high-pass filtered at 1\,Hz to allow sufficient averaging for fluctuations above this frequency, but eliminate contamination from lower-frequency oscillations. 
\textcolor{black}{This filtering means that any form of energy transfer in modes below 1 Hz is not measured by the technique. Together with the finite time resolution of the data discussed earlier, this means that the method is sensitive only to energy transfer in a specific range of spacecraft-frame frequencies, corresponding to $2\lesssim k\rho_\mathrm{i}\lesssim34$ under the Taylor hypothesis, which covers the majority of the kinetic range between the ion and electron gyroscales.}
In addition, the fluctuating distribution $\delta f_\mathrm{e}=f_\mathrm{e}-f_\mathrm{0e}$ was used, which removes the constant velocity-space structure that does not contribute to the small-scale energy transfer. The result is shown in Fig.~\ref{fig:2d}(c). It can be seen that the peak is more than an order of magnitude smaller, as expected for the secular transfer, and a qualitatively different pattern emerges: a symmetric pair of bipolar signatures at the thermal speed, evocative of Landau damping. \textcolor{black}{As discussed in the Methods section, other mechanisms would produce a qualitatively different signature.}

\begin{figure}
\includegraphics[width=\columnwidth,trim=0 0 0 0,clip]{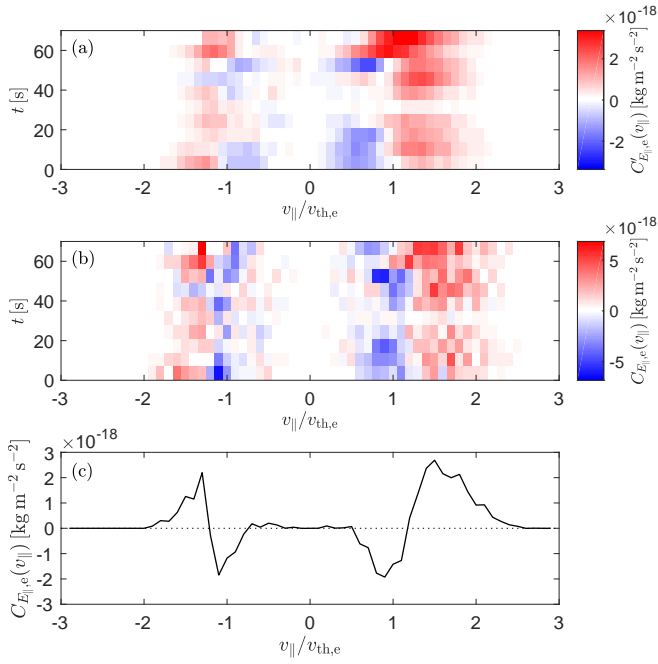}
\caption{\textbf{Reduced energy transfer rate measurements.} (a) Alternative 1D energy transfer rate $C'_{E_\mathrm{\|,e}}(v_\|)$ as a function of time $t$. (b) 1D energy transfer rate $C_{E_\mathrm{\|,e}}(v_\|)$ obtained from Eq.~(\ref{eq:C}). (c) $C_{E_\mathrm{\|,e}}(v_\|)$ averaged over time; a signature consistent with Landau damping can be seen.}
\label{fig:1d}
\end{figure}

To check whether this signature is coherent over time (which it should be for secular transfer and not for oscillatory transfer), the period was divided into 10 sub-intervals and the same analysis applied to each. Since the structure is mainly in $v_\|$, a reduced energy transfer measure was calculated, $C'_{E_\|,\mathrm{e}}(v_\|)=\int C'_{E_\|,\mathrm{e}}(\mathbf{v})\,\mathrm{d}^2\mathbf{v}_\perp$, which is shown in Fig.~\ref{fig:1d}(a) as a function of time. Due to the significant amount of averaging resulting in a less noisy signal, this could now be converted to the energy transfer rate specified in Eq.~(\ref{eq:cepar}) using the relation
\begin{equation}
C_{E_\mathrm{\|,e}}(v_\|)=-\frac{v_\|}{2}\frac{\partial C'_{E_\mathrm{\|,e}}(v_\|)}{\partial v_\|}+\frac{C'_{E_\mathrm{\|,e}}(v_\|)}{2},
\label{eq:C}
\end{equation}
which is shown in Fig.~\ref{fig:1d}(b). It can be seen that the symmetric bipolar pattern is indeed coherent over time, consistent with secular energy transfer to the electrons. The time average is shown in Fig.~\ref{fig:1d}(c), where the signatures consistent with electron Landau damping are present at velocities $\sim\pm v_\mathrm{th,e}$.

Finally, the curve in Fig.~\ref{fig:1d}(c) was integrated over $v_\|$ to obtain the net rate of secular transfer of energy density to the electrons $C_{E_\|,\mathrm{e}}\approx3.4\times10^{-12}\,\mathrm{kg}\,\mathrm{m}^{-1}\,\mathrm{s}^{-3}$. Comparing this to the electron thermal energy density, $\frac{3}{2}n_\mathrm{e}k_\mathrm{B}T_\mathrm{e}\approx 7.7\times10^{-11}\,\mathrm{kg}\,\mathrm{m}^{-1}\,\mathrm{s}^{-2}$, gives a transfer timescale of 23\,s, and comparing to the total thermal energy density, which is ten times larger, gives 230\,s. \textcolor{black}{This value of of $C_{E_\|,\mathrm{e}}$ is 6 times larger than the equivalent perpendicular quantity, $C_{E_\perp,\mathrm{e}}$, indicating a dominant parallel energy transfer to electrons in this interval.} It can also be compared to previously computed magnetosheath turbulent cascade rates \citep{hadid18}, where a wide variation of cascade rates were reported there in the range $\sim[10^{-16},10^{-12}]\,\mathrm{kg}\,\mathrm{m}^{-1}\,\mathrm{s}^{-3}$. The value of $C_{E_\|,\mathrm{e}}$ obtained here is at the upper end of this range, \textcolor{black}{consistent with the turbulence amplitude here being comparable to the upper end of the range of amplitudes \citep{hadid18}. This raises} the possibility that a significant fraction of turbulent energy is being transferred to electrons at kinetic scales.

\begin{figure}
\includegraphics[width=\columnwidth,trim=0 0 0 0,clip]{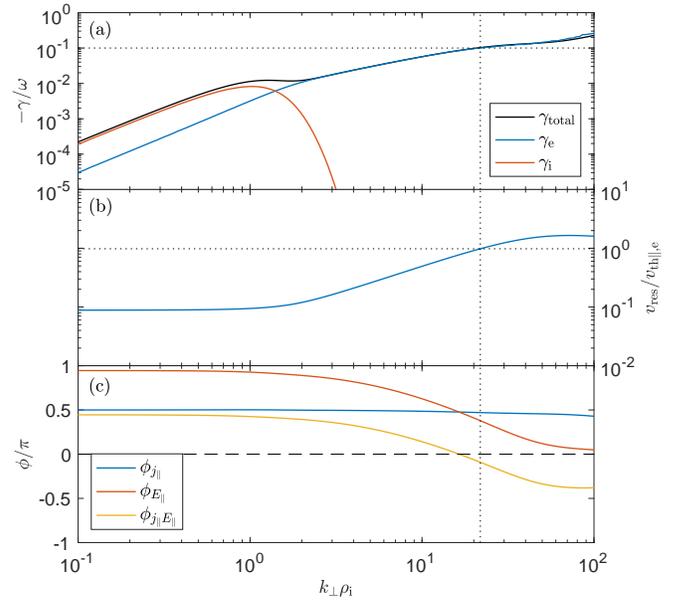}
\caption{\textbf{Numerical linear kinetic Alfv\'en wave solutions.} (a) Damping rates, (b) resonant velocity, and (c) phase angles for the KAW for the measured parameters. The dotted lines mark where the damping becomes strong ($-\gamma/\omega\sim0.1$).}
\label{fig:theory}
\end{figure}

\textbf{Comparison to expected resonant velocity.} The question now arises as to whether this signal occurs at the velocity expected for Landau damping. To answer this, Fig.~\ref{fig:theory} shows numerical solutions of the linear Vlasov-Maxwell system for the kinetic Alfv\'en wave (KAW) obtained from the PLUME dispersion solver \citep{klein15b}. The measured mean plasma parameters were used, along with $k_\|\rho_\mathrm{i}=10^{-3}$ (the results are not very sensitive to this number as long as it is $\ll1$). \textcolor{black}{Previous analysis of the data interval has suggested that the kinetic range fluctuations are low-frequency ($\omega\ll k_\perp v_\mathrm{th,i}$) and anisotropic ($k_\perp\gg k_\|$), consistent with kinetic Alfv\'en turbulence \citep{chen17}.} Fig.~\ref{fig:theory}(a) shows the total KAW damping rate normalised to the wave frequency, $-\gamma/\omega$, along with its separate contributions from the ions and electrons. It can be seen that the electron damping becomes strong ($-\gamma/\omega\sim0.1$) at a wavenumber $k_\perp\rho_\mathrm{i}\sim20$. Fig.~\ref{fig:theory}(b) shows the resonant velocity ($v_\mathrm{res}=\omega/k_\|$), which can be seen to be $v_\mathrm{res}\sim v_\mathrm{th\|,e}$ at $k_\perp\rho_\mathrm{i}\sim20$. Therefore, the locations of the energy transfer in Fig.~\ref{fig:1d} are consistent with expectations for Landau damping.

\begin{figure}
\includegraphics[width=\columnwidth,trim=0 0 0 0,clip]{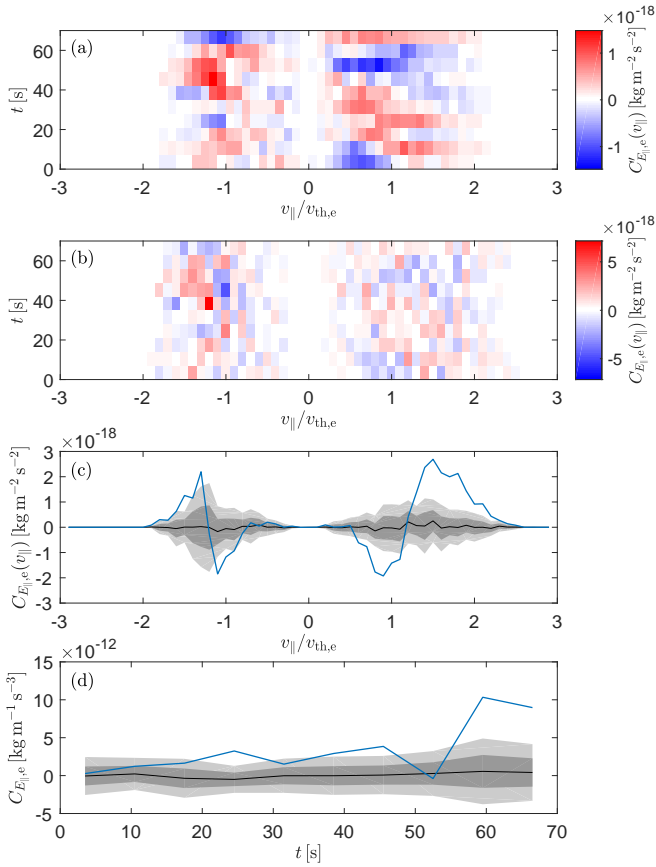}
\caption{\textbf{Reduced energy transfer rate measurements using a phase-randomised electric field.} (a-b) Same as in Fig.~\ref{fig:1d} except with randomised $E_\|$. (c) Energy transfer rate $C_{E_\|,\mathrm{e}}(v_\|)$ (blue) with the mean of the ensemble of phase randomisations (black) and shaded areas representing $\pm1\sigma$ (dark grey) and $\pm2\sigma$ (light grey). (d) Same for $C_{E_\|,\mathrm{e}}$ (integrated over $v_\|$) shown as a function of time $t$.}
\label{fig:phaseran}
\end{figure}

\textbf{Quality checks.} Several checks were performed to ensure that the field-particle correlation technique produced a meaningful result. Firstly, the analysis was repeated, but with \textcolor{black}{a phase-randomised version of the electric field measurement. To produce this, the electric field time series was Fourier transformed, a different random value chosen uniformly in the range $[0,2\pi]$ was added to the phase at each frequency, and then the inverse Fourier transform was applied. This results in a surrogate electric field time series with the same power spectrum and autocorrelation properties as the original \citep{prichard94}.} The results of the method using one realisation of the phase randomisation are shown in Fig.~\ref{fig:phaseran}(a-b). It can be seen that the pattern is quite different to Fig.~\ref{fig:1d}(a-b): the bipolar signatures are not present and the signal is not coherent over time. This suggests that the signals presented in Figs.~\ref{fig:2d} and \ref{fig:1d} are indeed physical. An ensemble of 20 realisations of the phase randomisation were performed and the mean and standard deviation $\sigma$ of the resulting $C_{E_\|,\mathrm{e}}(v_\|)$ are shown in Fig.~\ref{fig:phaseran}(c). \textcolor{black}{This number of realisations was chosen to allow sufficient convergence of the derived statistical quantities.} The mean is close to zero as expected and the amplitude of the real signal is large compared to the standard deviation, $\sim2\sigma$ for $v_\|<0$ and $\sim4\sigma$ for $v_\|>0$. Fig.~\ref{fig:theory}(c) shows that the phase angle between $j_\|$ ($\approx j_{\|\mathrm{e}}$ at these scales) and $E_\|$ is close to zero where the electron damping becomes strong, so indeed we would expect the phase randomisation to produce, on average, a smaller signal. Fig.~\ref{fig:phaseran}(d) shows $C_{E_\|,\mathrm{e}}=\int C_{E_\parallel,\mathrm{e}}(v_\parallel)\,\mathrm{d}v_\|=\left<j_{\|\mathrm{e}}E_\|\right>$ as a function of time in comparison to the phase randomisations. It can be seen that the real signal is consistently positive (indicating net energy transfer to the particles), whereas the phase randomisations are distributed about zero.

\begin{figure}
\includegraphics[width=\columnwidth,trim=0 0 0 0,clip]{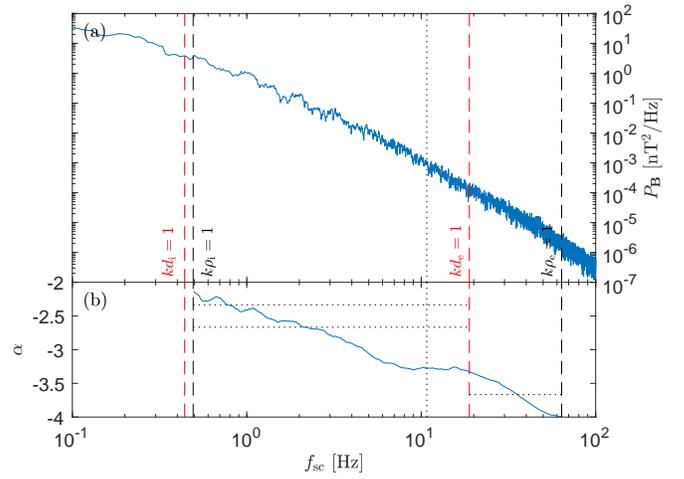}
\caption{\textbf{Magnetic field properties for the data interval.} (a) Trace magnetic field power spectrum $P_\mathbf{B}$ and (b) local spectral index $\alpha$; horizontal dotted lines mark asymptotic cascade predictions $-7/3$ \citep{schekochihin09} and $-8/3$ \citep{boldyrev12b} for kinetic Alfv\'en turbulence and $-11/3$ for inertial kinetic Alfv\'en turbulence \citep{chen17,passot17}, the vertical dotted line marks where the damping becomes strong.}
\label{fig:spectrum}
\end{figure}

The existence of Landau damping as the cascade proceeds towards electron scales can also be checked against the magnetic field spectrum; if energy is being removed from the turbulence, a steeper spectrum should result. The spectrum of magnetic fluctuations $P_\mathbf{B}$, and its local power-law index $\alpha$ (calculated over a sliding window of one decade) are shown in Fig.~\ref{fig:spectrum}. \textcolor{black}{If the turbulence is sufficiently low-frequency, which would be consistent with previous analysis \citep{chen17}, the Taylor hypothesis can be used to interpret this frequency spectrum as a wavenumber spectrum.} In the first decade of the kinetic range, $\alpha$ is comparable to predictions for kinetic Alfv\'en turbulence ($-7/3$ for a regular cascade \citep{schekochihin09} and $-8/3$ for an intermittent one \citep{boldyrev12b}), but by $k\rho_\mathrm{i}\approx15$ ($kd_\mathrm{e}\approx0.4$; $k\rho_\mathrm{e}\approx0.1$) it has steepened to a value of $-3.3$. This is significantly steeper than any current prediction for a dissipation-free cascade at these scales, consistent with a damping mechanism being in operation. Finally, as the cascade passes through the electron inertial scale $kd_\mathrm{e}=1$, the spectrum steepens again, consistent with expectations for an inertial kinetic Alfv\'en turbulence cascade \cite{chen17,passot17}. \textcolor{black}{Note also that for most of the frequency range, $\alpha$ is gradually decreasing rather than constant; while this is partly due to the finite width of the sliding window and the smallness of the frequency ranges, it is also consistent with the presence of damping progressively steepening the spectrum.}

\section*{Discussion}

The results presented in this paper constitute direct evidence for the presence of Landau damping in a turbulent space plasma, and suggest that it plays a significant role in the dissipation process. The secular energy transfer from the electric field to the electrons has been isolated from the oscillatory component (which has been measured previously in a KAW \citep{gershman17short}) and the electron distribution is seen to be gaining energy density above the resonant velocity and losing it below, with an overall net gain (Fig.~\ref{fig:1d}). The resonance appears rather broad, \textcolor{black}{with the signal covering a width comparable to or greater than the thermal speed (rather than a small fraction of it)}, as might be expected in strong turbulence \citep{quataert98,lynn12}. The observed velocity-space signature is consistent with simulations of Landau damping in both a single kinetic Alfv\'en wave and strongly nonlinear kinetic Alfv\'en turbulence \citep{klein17a}, and in the absence of other processes would correspond to a flattening of the distribution at the resonant velocity. However, additional processes, e.g. a velocity-space cascade \citep{schekochihin09,howes11a,told15} and/or collisions \citep{schekochihin09,borovsky11b}, would act to thermalise the distribution, so such plateaus may not be observable in practice.

An important question for understanding the kinetic turbulence itself, is the degree to which Landau damping steepens the energy spectrum \citep{howes08a,howes11a,howes11c,boldyrev12b,franci15b,told15,groselj17}. While fully answering this is beyond the scope of the current work, we note that the magnetic spectrum is significantly steeper than the cascade predictions at the scale where damping becomes strong (Fig.~\ref{fig:spectrum}). It is also of interest to note that the energy transfer is not uniform, fluctuating significantly in magnitude (Fig.~\ref{fig:phaseran}d) but maintaining a Landau-like signature (Fig.~\ref{fig:1d}b). This is consistent with suggestions that Landau damping is stronger at turbulent structures \citep{tenbarge13a,howes18}, and that in general dissipation in turbulence is intermittent in nature \citep{frisch95,wan12a,zhdankin14,zhdankin16a}.

Finally, we note that the results of the application of this technique are promising for its use in identifying the processes involved in turbulent dissipation. Although here we have determined the parallel energy transfer to electrons, future work could explore the perpendicular transfer and also the transfer to ions. With sufficiently advanced instrumentation on future spacecraft, this would allow the relative importance of the different mechanisms to be understood, as well as the energy partition between species and  the route by which heating is achieved in space and astrophysical plasmas.

\section*{Methods}

\textbf{Field-particle correlation technique.} The method for measuring the energy transfer is based on a field-particle correlation technique \citep{klein16b,howes17a,klein17b,klein17a,howes18} and briefly summarised here. The Vlasov equation,
\begin{equation}
\frac{\partial f_s}{\partial t}+\mathbf{v}\cdot\nabla f_s+\frac{q_s}{m_s}\left(\mathbf{E}+\mathbf{v}\times\mathbf{B}\right)\cdot\frac{\partial f_s}{\partial\mathbf{v}}=0,
\label{eq:vlasov}
\end{equation}
describes the evolution of the particle distribution function $f_s$ in a collisionless plasma, where $q_s$ and $m_s$ are the charge and mass of species $s$, $\mathbf{v}$ is the velocity, and $\mathbf{E}$ and $\mathbf{B}$ are the electric and magnetic fields. Multiplying by the particle kinetic energy, an equation for the rate of change of phase-space particle energy density $w_s=\frac{1}{2}m_sv^2f_s$ is obtained,
\begin{equation}
\frac{\partial w_s}{\partial t}=-\mathbf{v}\cdot\nabla w_s-\frac{q_sv^2}{2}\mathbf{E}\cdot\frac{\partial f_s}{\partial\mathbf{v}}-\frac{q_sv^2}{2}\left(\mathbf{v}\times\mathbf{B}\right)\cdot\frac{\partial f_s}{\partial\mathbf{v}}.
\label{eq:dws}
\end{equation}
When integrated over both position and velocity, only the second term on the right hand side of Eq.~(\ref{eq:dws}) is non-zero, showing that any net change in the particle energy is due to the electric field. This term has contributions from all electric field components, however here we focus on the energy transfer parallel to the magnetic field associated with Landau damping.

The average rate of change of phase-space energy density for species $s$ due to the parallel electric field $E_\|$ is given by
\begin{equation}
C_{E_\|,s}(\mathbf{v})=\left<-\frac{q_sv_\|^2}{2}E_\|\frac{\partial f_s}{\partial v_\|}\right>,
\label{eq:cepar}
\end{equation}
where the angular brackets denote an average over space and/or time. It can be seen that this is effectively an un-normalised correlation between $E_\|$ and the parallel gradient of the distribution function. Since such gradients are challenging to measure, we also define an alternative correlation,
\begin{equation}
C'_{E_\|,s}(\mathbf{v})=\left<q_sv_\|E_\|f_s\right>.
\label{eq:cepardash}
\end{equation}
When integrated over velocity space, Eqns.~(\ref{eq:cepar}) and (\ref{eq:cepardash}) are equivalent and correspond to the average net electromagnetic work done on the particles by $E_\|$,
\begin{equation}
\int C_{E_\|,s}(\mathbf{v})\,\mathrm{d}^3\mathbf{v}=\int C'_{E_\|,s}(\mathbf{v})\,\mathrm{d}^3\mathbf{v}=\left<j_{\|s}E_\|\right>,
\end{equation}
where $j_{\|s}$ is the parallel current density of species $s$.

An important part of the technique is the separation of the oscillatory transfer of energy back and forth between particles and fields due to undamped wave-like motions and the secular transfer due to damping (or instability). This is achieved by ensuring that the averaging period is much larger than the relevant wavelength and/or wave period.

In their unintegrated form, these correlation measures provide the crucial information about where in velocity space the secular energy transfer is occurring. Their application to simulations has shown that: (a) the oscillatory transfer can be successfully averaged out to leave the secular transfer, (b) a bipolar signature at the resonant velocity is produced for Landau damping of a single wave, (c) a qualitatively similar signature persists in strong low-frequency turbulence, and (d) the alternative measure [Eq.~(\ref{eq:cepardash})] indicates where in velocity space the transfer happens, although with a different characteristic signature \citep{klein16b,howes17a,klein17a}. \textcolor{black}{Energy transfer mechanisms other than Landau damping are expected to produce significantly different correlation signatures, e.g., cyclotron damping and stochastic heating would appear as perpendicular structure in the perpendicular correlations. Therefore, this technique allows the different mechanisms to be distinguished observationally.}

\section*{Data Availability}

The data used for this study is available at the MMS Science Data Center ({https://lasp.colorado.edu/mms/sdc/}).

\section*{Acknowledgments}

C.H.K.C is supported by STFC Ernest Rutherford Fellowship ST/N003748/2. K.G.K is supported by NASA HSR grant NNX16AM23G. G.G.H is supported by NSF CAREER Award AGS-1054061, NASA HGI grant 80NSSC18K0643, and NASA MMS GI grant 80NSSC18K1371. We thank the \emph{MMS} team for producing the data.

\section*{Author Contributions}

C.H.K.C performed the spacecraft data analysis and interpretation, wrote the initial manuscript text, and contributed to the development of the analysis method. K.G.K performed the numerical linear analysis and contributed to the development of the analysis method, data interpretation and manuscript text. G.G.H contributed to the development of the analysis method, data interpretation and manuscript text.

\bibliography{bibliography}

\end{document}